\documentclass[aip,reprint]{revtex4-1}
\usepackage{graphicx}
\usepackage{epstopdf}
\usepackage{booktabs}
\usepackage{grffile}
\usepackage{amsmath}

\begin{document}


\title{Adaptive motion mapping in pancreatic SBRT patients using Fourier transforms} 



\author{Bernard L. Jones}
\altaffiliation{Author to whom correspondence should be addressed: bernard.jones@ucdenver.edu}
\affiliation{Department of Radiation Oncology, University of Colorado School of Medicine}
\author{Tracey Schefter}
\author{Moyed Miften}
\affiliation{Department of Radiation Oncology, University of Colorado School of Medicine}


\newcommand{\super}[1]{\ensuremath{^{\textrm{{\tiny #1}}}}}
\newcommand{\subsc}[1]{\ensuremath{_{\textrm{{\tiny #1}}}}}
\newcommand{\enm}[1]{\ensuremath{#1}}
\newcommand{\mat}[1]{\overline{\overline{#1}}}

\begin{abstract}
\textbf{Purpose:} Recent studies suggest that 4DCT is unable to accurately measure respiratory-induced pancreatic tumor motion.   In this work, we assessed the daily motion of pancreatic tumors treated with SBRT, and developed adaptive strategies to predict and account for this motion.

\textbf{Methods}: The daily motion trajectory of pancreatic tumors during CBCT acquisition was calculated using a model which reconstructs the instantaneous 3D position in each 2D CBCT projection image.  We developed a metric (termed “Spectral Coherence,” SC) based on the Fourier frequency spectrum of motion in the SI direction, and analyzed the ability of SC to predict motion-based errors and classify patients according to motion characteristics.

\textbf{Results}: The amplitude of daily motion exceeded the predictions of pre-treatment 4DCT imaging by an average of 3.0 mm, 2.3 mm, and 3.5 mm in the AP/LR/SI directions.  SC was correlated with daily motion differences and tumor dose coverage.  In a simulated adaptive protocol, target margins were adjusted based on SC, resulting in significant increases in mean target D95, D99, and minimum dose.

\textbf{Conclusions}: Our Fourier-based approach differentiates between consistent and inconsistent motion characteristics of respiration and correlates with daily motion deviations from pre-treatment 4DCT. The feasibility of an SC-based adaptive protocol was demonstrated, and this patient-specific respiratory information was used to improve target dosimetry by expanding coverage in inconsistent breathers while shrinking treatment volumes in consistent breathers.

\textit{This manuscript was submitted to Radiotherapy and Oncology}

\end{abstract}


\maketitle 



%
%

\section{Introduction}

During treatment for pancreatic cancer, local failure is a major component of disease progression (30\%-50\%)\super{1}. To that end, clinicians have begun to pursue intensified radiotherapy regimens to improve patient outcomes\super{2-7}. Stereotactic Body Radiation Therapy (SBRT)\super{8,9} is especially appealing as a means of intensifying local therapy, and in recent years, encouraging clinical outcomes after SBRT for pancreas cancer have been reported from numerous centers\super{4-7,10}.  

SBRT relies on accurate delivery of ablative, conformal doses to the target; however, dose escalation in pancreatic cancer is complicated by the respiratory-induced motion of the abdomen.  The pancreas undergoes significant breathing-induced motion as a result of respiratory activity (as high as 2-3 cm), which necessitates expansion of the treatment margins\super{11-13}.  Additionally, the pancreas resides in close proximity to the duodenum and stomach, which are radiosensitive organs that, if damaged, can result in significant gastrointestinal (GI) toxicity\super{5,6}.  Under these conditions, pancreatic SBRT must balance a tradeoff between ensuring coverage of the mobile tumor while still limiting dose to nearby sensitive organs.

Compensation for respiratory motion is often accomplished using 4-Dimensional Computed Tomography (4DCT)\super{14}. By acquiring a series of 3D anatomical images throughout the respiratory cycle, 4DCT allows for the generation of a treatment volume which encompasses the entire range of motion of the tumor.  4DCT is used extensively to account for motion in treatment of thoracic and abdominal tumors, and generates accurate and reproducible data concerning lung and liver tumor motion\super{15,16}. Pancreatic tumors, on the other hand, tend to exhibit respiratory motion with more unstable, inconsistent qualities\super{17,18}.  As a result, recent studies have demonstrated discrepancies between pancreatic motion predicted by 4DCT and motion observed during treatment.  Minn \textit{et al} demonstrated that, in the majority of patients, pancreatic tumors could be found outside the bounds set by 4DCT over 10\% of the time during treatment\super{12}, and Ge \textit{et al} found that 4DCT underestimated the safety margin needed to account for superior-inferior motion by an average of 4 mm\super{13}.  This lack of an accurate tool to measure and compensate for patient-specific pancreatic motion further hinders the ability to escalate dose, as additional safety margins must be added to the target volume to account for these uncertainties. 

The aim of this study was to improve pancreatic SBRT by better understanding and accounting for motion.  To analyze the inaccuracies of 4DCT, the motion of pancreatic tumors was measured using a novel mathematical method based on projection imaging.  Based on this motion, we developed metrics which predict whether the motion observed during treatment will deviate from the motion observed during 4DCT imaging.  Finally, we investigated an adaptive treatment protocol which uses patient-specific respiratory information to derive individualized margins.

\section{Methods}
\subsection{Patients}
14 patients were treated for pancreatic carcinoma between October 2012 and February 2014.  Prior to treatment, CT simulation was performed to acquire a 3D CT for dose calculation and a 4-Dimensional CT (4DCT) for respiratory motion measurement. An abdominal bellows system was used to record the respiratory signal, and the acquired data were sorted into 10 phases based on temporal steps from maximal inspiration to another.  3D and 4DCT images were reconstructed with a slice thickness of 3 mm. In the 4DCT, the speed of table translation was set according to the patient’s breathing rate, such that each voxel in the image was under irradiation for one full breathing cycle.  Tumors were located either in the head (n=11) or body (n=3) of the pancreas, and 30 Gy was prescribed in 5 fractions to an internal target volume (ITV) which encompassed the range of motion observed on pre-treatment 4-Dimensional CT (4DCT).  Additionally, the ITV was expanded using patient-specific, 0-5 mm anisotropic margins to form the planning target volume (PTV).  For the purpose of this study, clinical tumor volumes (CTV) were also drawn on the 3D CT as part of our dosimetric analysis.  

To assist with localization, patients were implanted with radiopaque fiducial markers under endoscopic ultrasound guidance.  The cylindrical markers were approximately 5 mm long with a <1 mm diameter, and were composed either of gold (10 patients) or titanium with a carbon coating (4 patients).  Markers were implanted into the pancreatic periphery via trans-gastric insertions though duodenum, and during simulation and treatment, 1-6 markers (median=3) were present in the tumor. Patients were simulated and treated under free-breathing, and cone-beam CT (CBCT) was used to localize the target before treatment. In each patient, 5-7 CBCT images were acquired over the course of treatment (median=6).

\subsection{Tumor trajectory}
Pre-treatment 4DCT images were imported into the MIM Software system (MIM Software Inc, Cleveland, OH) for analysis.  Each image consisted of 10 respiratory phases; in each phase, the fiducial markers were contoured and the locations of the marker centroids were calculated.  These centroid positions were used to calculate the motion of the fiducial markers in the 4DCT images.  Figure 1 shows the approximate location of these tumors, and displays the motion of these tumors as observed in 4DCT images.

\begin{figure}
 \includegraphics{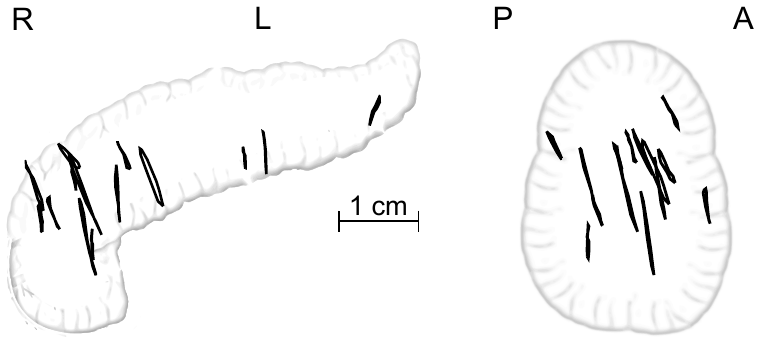}%
 \caption{Projected 3D trajectories of 14 pancreatic tumors in the coronal (left) and sagittal (right) planes, as measured from 4DCT images.  The trajectories show the movement of the tumor centroid, are drawn in approximate relation to one another using the common bile duct as positional reference. }
 \end{figure}

\begin{figure*}
 \includegraphics{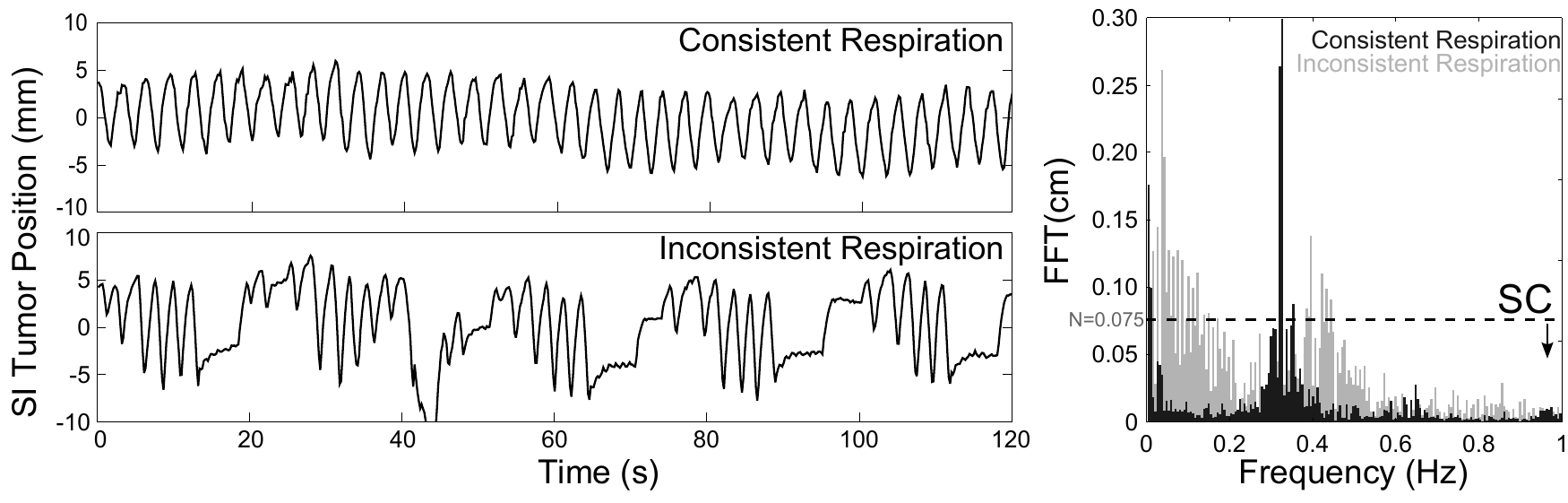}%
 \caption{Spectral Coherence and the consistency of respiration. The top left curve shows the tumor trajectory in the SI direction for a patient with consistent breathing, while the bottom left curve shows the trajectory of a patient with inconsistent breathing.  The Fourier Transform (FFT) of consistent motion is dominated by a single frequency, while the FFT of inconsistent motion contains contributions from many frequencies.  Spectral Coherence (right) is defined as the percentage of the motion spectrum below a given threshold value (denoted by the dashed line).  The SC is high (0.99) in the patient with consistent breathing, and low (0.95) in the patient with inconsistent breathing.}
 \end{figure*}

The daily trajectory of the tumor throughout the CBCT acquisition was calculated using a previously developed and validated method\super{19}. This technique was based on the work of Poulsen \textit{et a}l\super{20,21} and proceeded as follows.  CBCT images are formed from multiple 2D projection images acquired over time; in our method, each projection constitutes a measurement of the tumor position.  During CBCT acquisition, approximately 650 projection images were acquired over a 2 minute timeframe.  This method calculates the 3D position of the tumor (in patient geometry) using the position of the radiopaque fiducial markers in these projection images, and proceeds as follows.  First, the image locations of the fiducial markers were determined automatically in each projection using a template matching algorithm.  The location of the fiducial in the projection images confines the marker to lie along a line connecting the x-ray source to the point of detection in the imager.  This measurement was also corrected for the effect of gantry sag.  Next, a 3D Gaussian probability distribution function was computed for the daily tumor position distribution over all 650 images.  This distribution was calculated through maximum-likelihood optimization in order to best fit the observed locations measured in the projection images.  Finally, for each sampled position, the position of the marker was calculated in the patient by finding the most likely position along the line of detection (according to the Gaussian PDF).  Since pancreatic tumors are highly mobile (due to respiration), this analysis was performed independently for each phase of the breathing cycle. Respiratory phases were determined using the superior-inferior position of the fiducial markers in the projection images, and projections were sorted using temporal phase-binning. This model was implemented using an in-house MATLAB program (The Mathworks, Natick, MA).  

By applying this model, the daily trajectory of the tumor for each fraction of treatment was calculated.  A total of 81 CBCT datasets were analyzed, each containing roughly 650 projections acquired over 2 minutes.  The fiducial marker positions determined by the automated template matching algorithm were manually reviewed to ensure accuracy, and errors greater than two pixels were corrected (the position was determined correctly in roughly 95\% of the projections).  Using our method, each CBCT projection constituted a measurement of the tumor position in patient geometry (given by coordinates in the anterior-posterior, left-right, and superior-inferior directions, AP/LR/SI), otherwise known as the “tumor trajectory.”  The tumor position was measured roughly five times per second, and a full dataset encompassed 30-50 respiratory cycles.  Note that this is substantially more than the 4DCT, where each voxel contains motion information from only one respiratory cycle.

\subsection{Respiratory consistency}
Fourier spectrum analysis was used to understand and measure the consistency of respiration between patients.  For each tumor trajectory, the Fourier transform of tumor motion in the SI direction over time was calculated.  Figure 2 shows these trajectory data for a “consistent” and “inconsistent” breather.  Note that the spectrum of the consistent breather is dominated by a single frequency (~0.3 Hz, or 18 breaths per minute), whereas in the inconsistent breather, many frequencies ultimately contribute to the motion.  We propose to use a metric, Spectral Coherence, to describe this behavior.  Given some threshold value N, Spectral Coherence (SC) is equal to the fraction of the spectrum with Fourier transform values below N.  As SC approaches 100\%, the respiratory spectrum is dominated by fewer and fewer frequencies.  Likewise, lower values of SC indicate multiple respiratory frequencies contributing to tumor motion.  In this work, the discrete Fourier transform was calculated from 0 to 2.565 Hz in intervals of 0.005 Hz.

Each patient-specific trajectory exhibited at least some degree of respiratory inconsistency, and so the specific value of the threshold N determined what magnitude of consistency was considered meaningful.  In this study, the value of N was used which maximized the ability of SC to predict motion inconsistencies (i.e. maximize the correlation coefficient between SC and trajectory errors).  To cross-validate this value, bootstrap re-sampling was performed.  In this analysis, 10,000 random datasets were created by sampling (with replacement) from the full set of 81 tumor trajectories.  

\subsection{Motion metrics}
In clinical practice, 4DCT is often the gold standard for understanding and accounting for motion of the tumor during treatment.  To measure the effects of motion in our pancreatic patients, several metrics were calculated using the difference in expected position (from 4DCT) and actual position (using the tumor trajectory measured during CBCT acquisition).  The position of the tumor (either in 4DCT images or daily trajectories) was calculated as the mean position of all fiducial markers.  This analysis treats the tumor as a rigid volume, and assumes that the fiducial markers do not migrate significantly between simulation and treatment\super{22}.  To test this assumption, the differences in position between each marker were calculated in the planning CT and each CBCT.  No changes greater than 0.5 mm were observed, which excludes the presence of any migration or deformation greater than this distance and validates our assumptions of rigidity.  

The 4DCT trajectory described the motion of the tumor during ten phases of the respiratory cycle, and the daily trajectory described the daily motion of the tumor (650 samples over 2 minutes, across roughly 20-40 complete respiratory cycles).  After binning the daily trajectory into the same ten phases as the 4DCT, the error in position at each point in the trajectory was given as the absolute distance to the expected 4DCT position (for that phase).  These values were computed relative to the centroid of tumor motion.  By taking the mean absolute difference across the entire daily trajectory, the “Mean Trajectory Error” was calculated. This metric describes the average magnitude of the displacement between the 4DCT position and the actual position at a given moment in time.

\subsection{Dosimetric calculations}
To evaluate the dosimetry of a given tumor motion trajectory, a hypothetical clinical scenario was considered.  The clinically-drawn CTV was expanded using the observed 4DCT motion to form a simulated ITV.  This ITV was then expanded by some anisotropic margin to form a simulated PTV.  The CTV (a surrogate for the gross tumor) underwent the daily motion trajectory calculated from the CBCT projections; for each measured position in the trajectory, the position of the CTV relative to its center was given by the calculated tumor position relative to the mean position. Based on this translation, the position of each voxel in the CTV was calculated. The PTV was stationary with respect to the mean CTV position; in other words, as the CTV moved with respiration, the PTV remained stationary in absolute space. To simplify the model, and to exaggerate differences between different trajectories and margin strategies, the dose within the PTV was assumed to equal the prescription, with zero dose outside the PTV.  Based on the CTV location relative to the PTV, the dose was calculated for each voxel in the CTV.  

The predictive power of Spectral Coherence was investigated by retrospectively implementing an adaptive protocol to account for patient-specific motion inconsistencies.  The 4DCT-based CTV volumes were used as target volumes.  In the non-adaptive case, the ITV-to-PTV margin was equal to 3 mm in the AP/LR directions and 5 mm in the SI direction.  In the adaptive case, margins were adjusted for each patient according to SC.  In patients with consistent breathing, the margin for all fractions was lowered to 3 mm in the SI direction; in cases with less consistent breathing, the margin was increased to 7 mm in the SI direction.  Specifically, margins were expanded for patients in the lower quartile of SC, contracted in the upper quartile, and remained the same in the middle 50\%.  Dose-volume histograms were computed to compare the results of the adaptive and non-adaptive protocol. Differences in the adaptive and non-adaptive protocols were compared using a paired T-test.

\begin{figure}
 \includegraphics{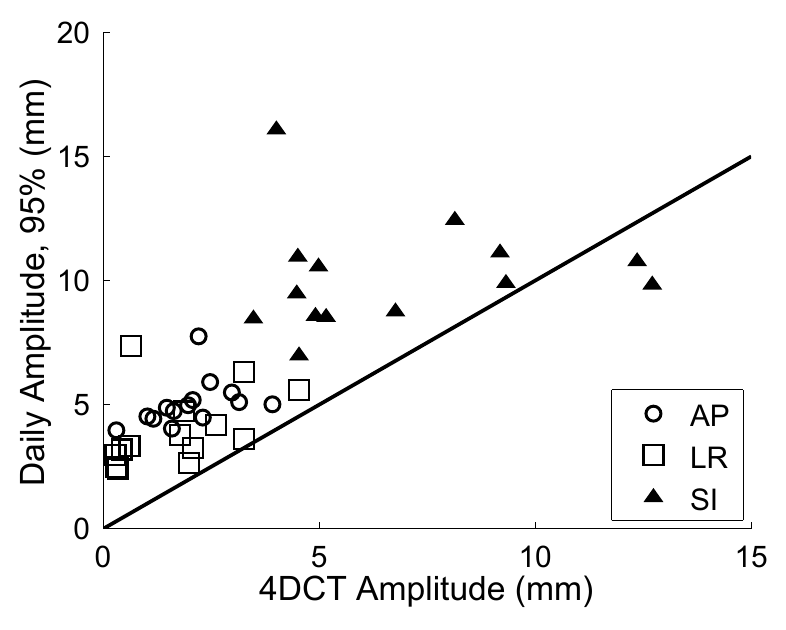}%
 \caption{Motion amplitude of pancreatic tumors in the anterior-posterior (AP), left-right (LR), and superior-inferior (SI) directions.  4DCT amplitudes were measured from pre-treatment 4DCT images.  Daily amplitudes are derived from the reconstructed tumor trajectories, and are shown here as the distance encompassing 95\% of the observed motion.   The solid line has a slope of 1, and denotes where the data would fall if the amplitudes agreed.  In nearly every case, 4DCT underestimated the amplitude of motion. In the SI direction, motion on the day of treatment exceeded the value predicted from 4DCT by an average of 3.5 mm.}
 \end{figure}

\begin{figure*}
 \includegraphics{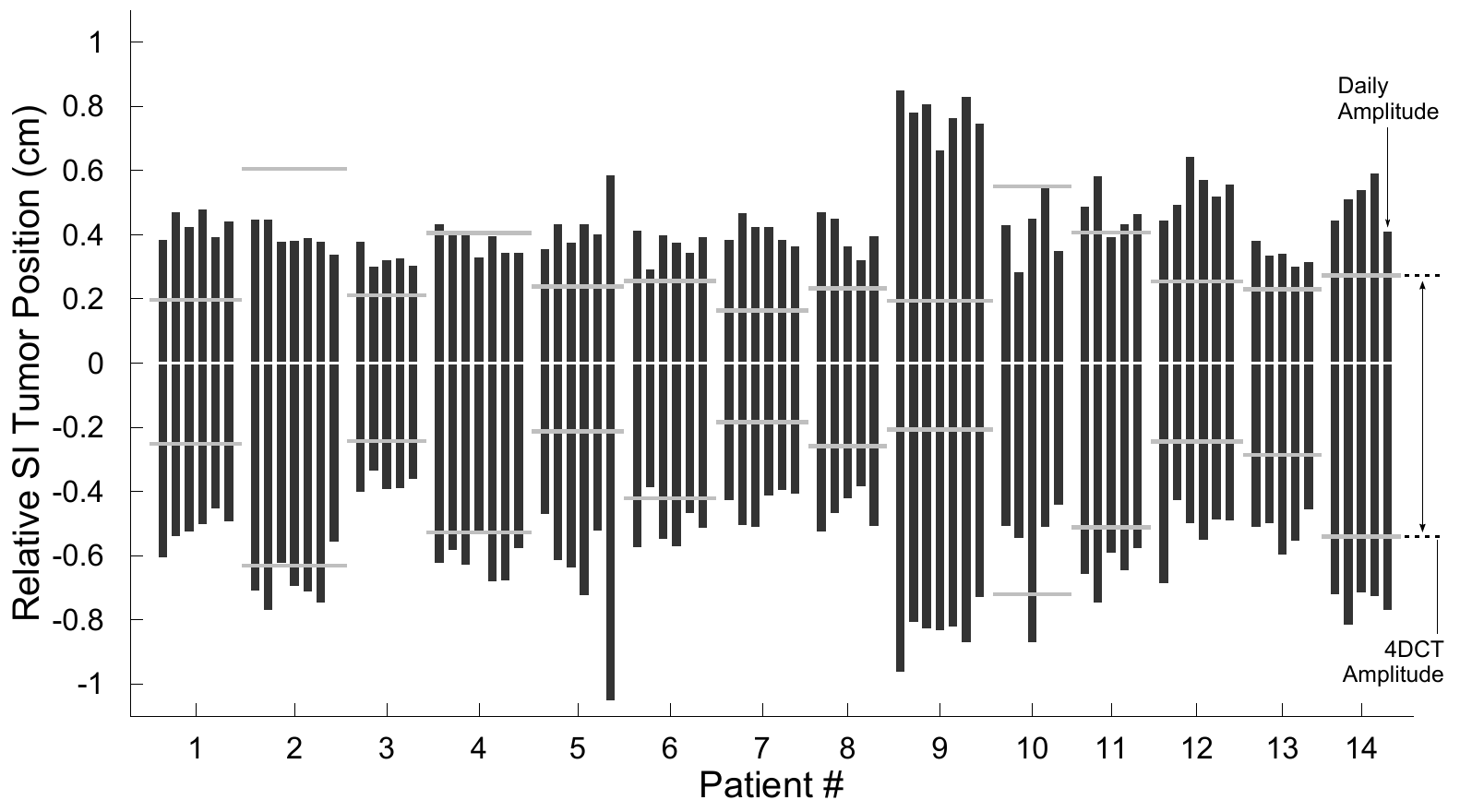}%
 \caption{Tumor trajectories in the superior-inferior (SI) direction. Each vertical box denotes the distance encompassing 95\% of the daily motion during one fraction of treatment.  Fractions are grouped by patient, and the horizontal bars denote the amplitude of SI motion measured in the 4DCT images for that patient.  In addition to highlighting the tendency of 4DCT to underestimate pancreatic motion, these data also demonstrate the relative stability of motion in a single patient (from day-to-day), as well as the large differences in motion between different patients.  Tumor position and 4DCT amplitude are shown relative to the mean position.}
 \end{figure*}

\section{Results}

\subsection{Inaccuracies of 4DCT}
Using the projection dataset from each CBCT acquisition, the daily trajectories of each tumor were reconstructed.  The daily motion amplitude (in the AP, LR, and SI directions) was computed from these trajectories.  To exclude transient breathing excursions and/or outliers, the daily amplitude was computed for each dataset as the distance encompassing 95\% of the measured positions, and was averaged for each patient across all fractions of treatment.  Figure 3 compares this daily amplitude to the predicted motion from 4DCT imaging.  4DCT significantly underestimated the amplitude of motion (p < 0.01), and the mean differences from the measured daily amplitude were 3.0 mm (AP), 2.3 mm (LR), and 3.5 mm (SI).

\subsection{Consistency of Tumor Trajectories}
Figure 4 shows the tumor trajectories in the SI direction for all fractions of treatment in all patients.    First, these data provide further confirmation as to the inaccuracies of 4DCT in predicting motion of pancreatic tumors.  Each dot represents one measurement of the tumor position, and each measurement beyond the 4DCT margins represents a moment when the observed daily motion exceeded the pre-treatment prediction.  Figure 4 also highlights the large differences in motion between different patients.  Moreover, these data demonstrate the consistency of motion within each individual patient; in particular, the relative stability of the uncertainty in SI tumor position (defined here as 1 standard deviation of the SI trajectory).  Across all datasets, the uncertainty of a single trajectory differed from that patient’s average uncertainty by 8.3\% \enm{\pm} 7.4\%.  

\begin{figure}
 \includegraphics{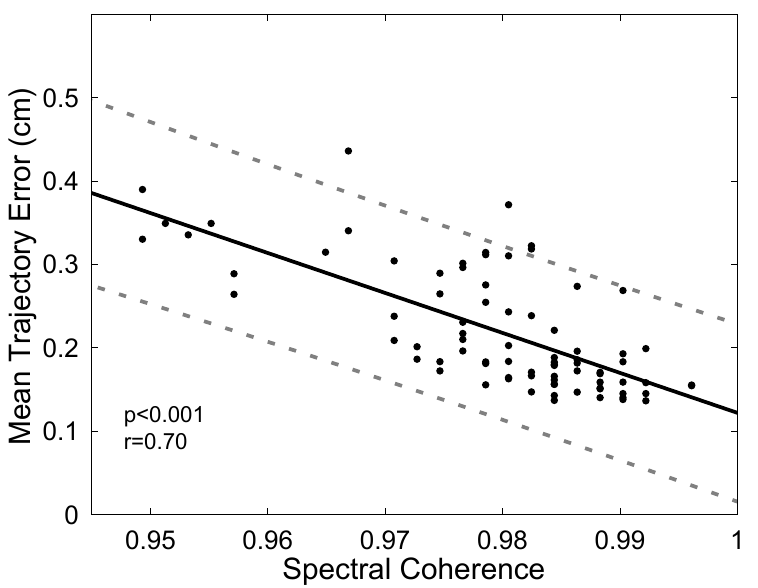}%
 \caption{Spectral coherence is highly correlated with mean trajectory error. The solid line denotes the least-squares regression between the two quantities, and the dotted lines denote the 95\% prediction interval for determining mean trajectory error from a given spectral coherence value.}
 \end{figure}

\subsection{Spectral Coherence}
The optimum value of the threshold N (for the calculation of SC) in our dataset was 0.075 cm, equal roughly to the 98th percentile of values in the Fourier transform (i.e. only 2\% of respiratory frequencies had amplitudes greater than this value in the Fourier analysis).  From bootstrap resampling analysis, the standard deviation in the optimum threshold was 0.016 cm, with a 95\% confidence interval of 0.029 – 0.086 cm. Across all trajectories, the mean SC value was 0.980 ± 0.011 (range, 0.949 – 0.996). The average difference between SC from a single fraction and that patient’s average SC was 0.004, whereas the standard deviation of mean SC between patients was 0.011.  In other words, the deviations in SC between patients were substantially larger than the deviations in SC in a single patient.  Note that since N is derived from the FFT, it depends on the sampling interval (5.48 Hz in this dataset).

Figure 5 shows the relationship between mean trajectory error and spectral coherence (SC).  These data demonstrate a strong correlation between SC (a measure of respiratory consistency) and mean trajectory error (which quantifies differences between 4DCT and observed daily motion).  SC explains a large fraction of the variability in mean trajectory error (R2 = 0.491), and the 95\% prediction interval for a given SC value is 2.1 mm.  In addition, it was seen that SC added predictive value over the more straightforward approach of using the magnitude of motion as a predictor of trajectory error.  The correlation coefficient between motion magnitude and mean trajectory error was r=0.55 (compared to r=0.70 for SC).   Additionally, there was less than 10\% deviation in correlation between SC and mean trajectory error across the 95\% confidence interval of the threshold N.

A strong correlation was also seen between SC and dosimetric quantities D90 (r=0.75), D95 (r=0.77), D99 (r=0.78), and minimum dose (r=0.79).  These dose metrics were calculated using the simplified clinical model which incorporates the measured motion trajectories (Sect \textit{Dosimetric Calculations}). 

\subsection{Adaptive Treatment}
Table 1 shows the results of the SC-based adaptive protocol.  In the adaptive scenario, ITV-to-PTV margins were expanded for patients with in the lower quartile of mean SC, and shrunk in trajectories in the upper quartile of mean SC.  D90, D95, and D99 were significantly increased under the adaptive protocol. Also, minimum target dose trended towards a significant increase.  In the adaptive case, the average margin was the same as in the non-adaptive case; however, by using SC to selectively add or spare margins based on the consistency of breathing, the treatments were improved.
 
\begin{table}[h]
\begin{tabular}{@{}lccc@{}}
\toprule
\hline
\multicolumn{1}{c}{}  & \textbf{Non-Adaptive} & \textbf{Adaptive} & \textit{\textbf{p}} \\ \midrule
\hline
\textbf{D90}          & 99.2\enm{\pm}2.6              & 99.7\enm{\pm}0.8          & 0.01                \\
\textbf{D95}          & 98.8\enm{\pm}3.1              & 99.4\enm{\pm}1.3          & 0.01                \\
\textbf{D99}          & 97.8\enm{\pm}4.9              & 98.3\enm{\pm}3.1          & 0.03                \\
\textbf{Minimum Dose} & 95.6\enm{\pm}7.4              & 96.0\enm{\pm}5.6          & 0.08                \\ \bottomrule
\hline
\end{tabular}
\caption{Adaptive treatment based on spectral coherence improves dosimetry.  In the non-adaptive case, CTV-to-PTV margins were 3 mm in the LR/AP directions, and 5 mm in the SI direction.  In the adaptive case, SI margins were decreased 2 mm for the upper quartile of SC, and increased 2 mm in the lower quartile of SC.}
\end{table}

\section{Discussion}

By combining high doses per fraction with small target volumes, SBRT can achieve large dose gradients between a tumor and the surrounding normal tissue.  However, accurate and conformal dose delivery is crucial to the success of SBRT, and motion of the tumor presents a challenge to this therapy by blurring the boundary between these structures.  In our patients, 4DCT was unable to accurately measure and account for this motion; on average, the motion of the tumor on the day of treatment exceeded the motion measured from 4DCT by 3.0 mm, 2.3 mm, and 3.5 mm in the AP, LR, and SI directions. These results agree well with previously published studies on pancreatic motion\super{12,13}, and serve to further confirm the drawbacks of 4DCT in abdominal imaging. The cause of these discrepancies is unknown; however, there are several plausible explanations.  4DCT uses external surrogates of motion (such as a chest block or bellows) to determine respiratory phase, and this motion may not correlate with the actual internal motion (as seen in this study).  Moreover, 4DCT images are formed from only a handful of respiratory cycles, and if breathing in a given patient is inconsistent from cycle to cycle then 4DCT data would be a poor representation of the motion over a longer time period.  Respiratory coaching may limit this effect.  In addition, it is possible that some patients may exhibit motion changes in the time interval between simulation and treatment; however, this is less likely given the relative stability of motion observed from day-to-day of treatment (Fig 4).  

We propose the use of Spectral Coherence (SC) to better account for patient-specific difference in pancreatic tumor motion, a metric which describes the relative contribution of different frequencies within the respiratory cycle to the overall motion of the tumor.  Patients with consistent breathing exhibit respiratory-induced tumor motion which is dominated by a single (coherent) frequency, whereas in patients with inconsistent breathing, many different frequencies of motion contribute significantly to the observed tumor path.  Daily respiratory inconsistencies are comprised of varying respiratory amplitudes, baseline shifts, and varying respiratory periods; although SC cannot differentiate between these different factors affecting motion inconsistencies, it distills information regarding this inconsistency into a single parameter, from which a clinical judgment can be made. Our results show that daily SC is highly correlated with deviations from 4DCT motion. Additionally, in a simplified clinical scenario to determine the dosimetric effect of that motion, SC was also highly correlated with D90, D95, D99, and minimum dose to the tumor.  Our calculations depend on the specific value N chosen as the threshold for computation of SC; in our analysis it was shown that the ability of SC to predict trajectory errors is stable across the 95\% confidence interval of the value for N (0.075 cm) used in this study.

These results also demonstrate the clinical relevance of SC.  In a hypothetical adaptive protocol, the average tumor dosimetry was improved by expanding and contracting the PTV margins in cases with low and high SC, respectively.  The utility of this approach is twofold. By identifying patients with consistent breathing and shrinking the size of treatment volumes in these cases, SC points towards cases where dose escalation may be beneficial.  Likewise, by identifying patients with inconsistent breathing, it is possible to design treatment volumes that better cover the full range of tumor motion.  

It remains to be seen how best to measure SC clinically.  In this study, SC was measured from trajectories derived from CBCT acquisitions, and in these cases no SC information is available until the first fraction of treatment.  However, within a single patient, it was observed that the motion remained stable throughout all fractions of treatment, and it was seen that the trajectory derived from a single day’s CBCT acquisition can predict the general characteristics of that patient’s future trajectories with an error of roughly 10\%.  For patients receiving three to five fractions of SBRT, this suggests that it may be possible to adapt the course of treatment based on respiratory motion observed during the initial fraction.  A better workflow would be to measure SC pre-treatment, and it is possible, in theory, to measure SC using fluoroscopic imaging at the time of simulation. However, this extra pre-treatment imaging may constitute an additional burden to the patient.

\section{Conclusions}

Using projection imaging, the daily motion of 14 pancreatic tumors undergoing SBRT was measured immediately prior to each fraction of treatment. The amplitude of this motion exceeded the predictions of pre-treatment 4DCT imaging by an average of 3.0 mm, 2.3 mm, and 3.5 mm in the AP, LR, and SI directions.  We developed a Fourier-based approach which differentiates between consistent and inconsistent motion characteristics of respiration using a quantity termed Spectral Coherence (SC).  Lower values of SC were strongly associated with larger deviations in motion from 4DCT predictions, and SC was positively correlated with improvements in target dosimetry.  The feasibility of an SC-based adaptive protocol was demonstrated, and this patient-specific respiratory information was used to improve target dosimetry by expanding coverage in inconsistent breathers while shrinking treatment volumes in consistent breathers.  

\section{Acknowledgements}
This work was funded in part by the National Institutes of Health under award number K12CA086913, and by an award from Varian Medical Systems. These funding sources had no involvement in the study design; in the collection, analysis and interpretation of data; in the writing of the manuscript; or in the decision to submit the manuscript for publication.

\section*{References}
{\small 
[1]	Willett CG, Czito BG, Bendell JC Ryan DP. Locally advanced pancreatic cancer. Journal of Clinical Oncology 2005;23:4538-4544.

[2]	Ceha HM, van Tienhoven G, Gouma DJ, et al. Feasibility and efficacy of high dose conformal radiotherapy for patients with locally advanced pancreatic carcinoma. Cancer 2000;89:2222-2229.

[3]	Koong AC, Le QT, Ho A, et al. Phase I study of stereotactic radiosurgery in patients with locally advanced pancreatic cancer. International Journal of Radiation Oncology* Biology* Physics 2004;58:1017-1021.

[4]	Chang DT, Schellenberg D, Shen J, et al. Stereotactic radiotherapy for unresectable adenocarcinoma of the pancreas. Cancer 2009;115:665-672.

[5]	Koong AC, Christofferson E, Le QT, et al. Phase II study to assess the efficacy of conventionally fractionated radiotherapy followed by a stereotactic radiosurgery boost in patients with locally advanced pancreatic cancer. International Journal of Radiation Oncology* Biology* Physics 2005;63:320-323.

[6]	Mahadevan A, Miksad R, Goldstein M, et al. Induction gemcitabine and stereotactic body radiotherapy for locally advanced nonmetastatic pancreas cancer. Int J Rad Oncol* Biol* Phys 2011;81:e615-e622.

[7]	Chuong MD, Springett GM, Freilich JM, et al. Stereotactic body radiation therapy for locally advanced and borderline resectable pancreatic cancer is effective and well tolerated. International Journal of Radiation Oncology* Biology* Physics 2013;86:516-522.

[8]	Kavanagh BD, Miften M Rabinovitch RA. Advances in Treatment Techniques: Stereotactic Body Radiation Therapy and the Spread of Hypofractionation. The Cancer Journal 2011;17:177-181.

[9]	Kavanagh BD Timmerman RD. Stereotactic radiosurgery and stereotactic body radiation therapy: an overview of technical considerations and clinical applications. Hematol Oncol Clin North Am 2006;20.

[10]	Rwigema JCM, Parikh SD, Heron DE, et al. Stereotactic body radiotherapy in the treatment of advanced adenocarcinoma of the pancreas. American Journal of Clinical Oncology 2011;34:63.

[11]	Mori S, Hara R, Yanagi T, et al. Four-dimensional measurement of intrafractional respiratory motion of pancreatic tumors using a 256 multi-slice CT scanner. Radiotherapy and Oncology 2009;92:231-237.

[12]	Minn AY, Schellenberg D, Maxim P, et al. Pancreatic tumor motion on a single planning 4D-CT does not correlate with intrafraction tumor motion during treatment. American Journal of Clinical Oncology 2009;32:364.

[13]	Ge J, Santanam L, Noel C Parikh PJ. Planning 4-Dimensional Computed Tomography (4DCT) Cannot Adequately Represent Daily Intrafractional Motion of Abdominal Tumors. Int J Rad Oncol* Biol* Phys 2012.

[14]	Pan T, Lee TY, Rietzel E Chen GTY. 4D-CT imaging of a volume influenced by respiratory motion on multi-slice CT. Medical Physics 2004;31:333.

[15]	Guckenberger M, Wilbert J, Meyer J, Baier K, Richter A Flentje M. Is a Single Respiratory Correlated 4D-CT Study Sufficient for Evaluation of Breathing Motion? International Journal of Radiation Oncology*Biology*Physics 2007;67:1352-1359.

[16]	Guckenberger M, Sweeney RA, Wilbert J, et al. Image-Guided Radiotherapy for Liver Cancer Using Respiratory-Correlated Computed Tomography and Cone-Beam Computed Tomography. IJROBP 2008;71:297-304.

[17]	Feng M, Balter JM, Normolle D, et al. Characterization of pancreatic tumor motion using cine MRI: surrogates for tumor position should be used with caution. International Journal of Radiation Oncology* Biology* Physics 2009;74:884-891.

[18]	Heerkens HD, van Vulpen M, van den Berg CA, et al. MRI-based tumor motion characterization and gating schemes for radiation therapy of pancreatic cancer. Radiotherapy and Oncology 2014.

[19]	Jones BL, Westerly D Miften M. Calculating tumor trajectory and dose-of-the-day using cone-beam CT projections. Medical Physics 2015;42:694-702.

[20]	Poulsen PR, Cho B Keall PJ. A method to estimate mean position, motion magnitude, motion correlation, and trajectory of a tumor from cone-beam CT projections for image-guided radiotherapy. International Journal of Radiation Oncology* Biology* Physics 2008;72:1587-1596.

[21]	Poulsen PR, Cho B Keall PJ. Real-time prostate trajectory estimation with a single imager in arc radiotherapy: A simulation study. Physics in medicine and biology 2009;54:4019.

[22]	Sanders MK, Moser AJ, Khalid A, et al. EUS-guided fiducial placement for stereotactic body radiotherapy in locally advanced and recurrent pancreatic cancer. Gastrointestinal endoscopy 2010;71:1178-1184.
}


\end{document}